\documentclass[english]{nature}
\usepackage[colorlinks=true,citecolor=blue,linkcolor=magenta]{hyperref}
\usepackage{lineno}
\usepackage{amsmath}
\usepackage{amsfonts}
\usepackage{stmaryrd}
\usepackage{amssymb}
\usepackage{graphicx}% Include figure files
\usepackage{dcolumn}% Align table columns on decimal point
\usepackage{bm}% bold math
\usepackage[font={sf, small},labelfont=bf]{caption}
\usepackage{siunitx}
\usepackage{soul}
\usepackage{changes}
\usepackage{indentfirst}
\usepackage{float} 
\usepackage{longtable}

\title{A high-resolution microresonator-frequency-comb spectrometer}
\author
{\noindent Ruocan Zhao$^{1, \dagger}$, Bin Yang$^{1, \dagger}$, Chuan Huang$^{1}$, Jiangtao Li$^{1}$, Baoqi Shi$^{2}$, Wei Sun$^{2}$, Chen Shen$^{2,3}$, Chong Wang$^{1}$, Tingdi Chen$^{1,4}$, Chen Liang$^{4}$, Xianghui Xue$^{1,4,5,6,\ast}$, Junqiu Liu$^{2,4}$, and Xiankang Dou$^{4}$}
\begin{document}
\maketitle
\begin{affiliations}
\item CAS Key Laboratory of Geospace Environment, School of Earth and Space Sciences, University of Science and Technology of China, Hefei 230026, China
\item International Quantum Academy, Shenzhen 518048, China
\item Qaleido Photonics, Shenzhen 518048, China
\item Hefei National Laboratory, University of Science and Technology of China, Hefei 230088, China
\item CAS Center for Excellence in Comparative Planetology, Anhui Mengcheng Geophysics National Observation and Research Station, University of Science and Technology of China, Hefei, China
\item  Hefei National Research Center for Physical Sciences at the Microscale and School of Physical Sciences, University of Science and Technology of China, Hefei, China.\\
\normalsize{$^\dagger$  These authors contributed equally to the work}\\
\normalsize{Emails: $^\ast$ xuexh@ustc.edu.cn}
\end{affiliations}

\begin{abstract}
Spectral analysis is one of the most powerful technologies for studying and understanding matter.
As the devices for spectral analysis, spectrometers are widely used in material detection, isotope analysis, trace gas detection, and the study of atomic and molecular hyperfine structures.
While high resolution, wide bandwidth and fast speed are essential factors, they are always trade-offs for conventional spectrometers.
Here, we present a soliton-microcomb-based spectrometer that overcomes these challenges by integrating dissipative Kerr solitons (DKSs) with double-sideband modulation and parallelized detection. 
Leveraging a high-quality silicon nitride microresonator, we generate a broadband, fully stabilized soliton microcomb and employ radio-frequency-modulated double sidebands to scan the optical spectrum with the resolution constrained only by the comb-line linewidth. 
By projecting the comb lines onto a two-dimensional charge-coupled device (CCD) via a virtually imaged phased array (VIPA)-grating system, we enable parallel processing of all spectral components, circumventing sequential scanning delays. 
The resulting spectrometer achieves 200-kHz resolution across a 4-THz bandwidth with minutes-level processing time while maintaining robustness against environmental fluctuations. 
Being promising for miniaturization, this work bridges the gap between laboratory-grade performance and field-deployable practicality, unlocking new possibilities for spectroscopy in astronomy, metrology, and integrated photonics.
\end{abstract}

%%%%%%%%%%%%%%%%%%%%%%%
%%%%% Introduction %%%%%
%%%%%%%%%%%%%%%%%%%%%%%

\section*{Introduction}

Matter ubiquitously interacts with light, which transfers matter's information onto the amplitude, phase, and frequency of the light.
Spectroscopy is such a pivotal branch of science aiming to obtain matter information, such as the physical structures, the chemical ingredients, and the dynamical states, by analyzing the spectra of the interacted light.

The devices for light spectra analysis are commonly referred to as optical spectrometers (or simply, spectrometers).
Spectrometers have long been investigated and applied across the fields of physics, chemistry, medicine, biology, astronomy, and material science.
The steps pursuing higher measurement resolution never halt over the last century, as well as in the future.
Types of conventional spectrometers include the diffraction grating spectrometers\cite{Palmer:05}, the Fourier spectrometers\cite{Griffiths:83} and the Fabry-P\'erot (FP) spectrometers\cite{Hernandez:88}, with measurement resolution down to sub-GHz level.
The invention of the optical frequency comb (OFC)\cite{Udem:02, Cundiff:03, Fortier:19} has revolutionized the field of spectroscopy. 
While allowing measurements with unparalleled precision\cite{Coddington:16, Picque:19, Diddams:20}, the measurement resolution of the OFC spectroscopy is limited to the comb mode spacing, typically above 100 MHz.
The other drawbacks of OFC spectroscopy include the small dynamical range because of the weak power of the individual comb line and the damaging risk on the photodetector because of the high total power. 
To circumvent the drawbacks and further improve the measurement resolution, continuous tunable lasers (CTLs) are introduced in combination with the OFC calibration\cite{DelHaye:09, Giorgetta:10, Shi:24}, the single-sideband modulation\cite{Tang:12, Pan:17}, the OFC-assisted asymmetric double-sidebands modulation\cite{Qing:19}, the Mach-Zender interferometers\cite{Li:12, Twayana:21} and the fiber-cavity calibration\cite{Luo:24}, which can tremendously improves the measurement resolution to the level of the laser frequency linewidth (Hz to sub-MHz depending on the laser performance).

Although endowed with excellent measurement resolution, high-resolution spectrometers based on CTLs are usually inconvenient for wide applications outside of the laboratories.
For the OFC calibration scheme, the OFCs generated from mode-locked lasers and the wideband CTLs are bulky and costly.
For the single-sideband modulation scheme, the bandwidth is shallow and limited to 100 GHz.
For the OFC-assisted asymmetric double-sideband modulation scheme, a wideband measurement could consume dozens of minutes even hours because the comb lines have to be processed in sequence.
For the Mach-Zender interferometers and the fiber cavities schemes, the two kinds of devices are environment dependent whose free spectral ranges (FSRs) can be influenced by temperature and vibrations.
In recent years, extensive efforts have been made to minimize the spectrometers with small size and weight\cite{YangZY:21, Li:22, Ryckeboer:13, Pohl:20, YangZY:19, Yoon:22, Zhang:22, Ni:22, LiYH:23, Cen:23, Toulouse:21, Xu:23b} while ubiquitously at the cost of low bandwidth or resolution.

The instant alternatives of the mode-locked lasers are the DKSs\cite{Herr:14, Kippenberg:18, Gaeta:19, Xue:15} generated in high-quality microresonators. 
Commonly referred to as soliton microcombs, they are featured with broad bandwidth and repetition rates in the microwave to millimeter-wave domain.
Various photonic integrated platforms have been developed for soliton microcomb generation\cite{Yi:15,Brasch:15, Joshi:16, Liang:15, Yang:18, He:19, Wang:19, Bao:19, LiuX:21,Pu:16, Guidry:21, Jung:21,XiaD:22, Yao:24}.
Meanwhile, soliton microcombs have already been used in many system-level information and metrology applications\cite{Marin-Palomo:17, Trocha:18, Suh:18, Obrzud:19, Suh:19, Dutt:18, Yang:19, Feldmann:21, Xu:21, Spencer:18, Newman:19, Shu:22}.
Hybrid and heterogeneous integration\cite{Stern:18, Xiang:21, Liu:20a, Churaev:23, OpdeBeeck:20, Sun:24} further enable complex control schemes, extra nonlinearity, and efficient amplification for integrated soliton microcombs.

Here we demonstrate a soliton-microcomb-based spectrometer with the assistance of the double-sidebands modulation.
By pumping a high-quality silicon nitride (Si$_3$N$_4$) microresonator with anomalous dispersion, a wideband soliton microcomb is generated.
The microcomb is fully locked (immune to the circumstance) and further modulated with every comb line split into double sidebands. 
By continuously changing the modulation frequency, the sidebands are scanned to fulfill the comb line spacing and transfer the sub-Hz precision of the radio frequency (RF) domain to the optical domain.
The resolution of the spectrometer is thereby limited by the comb line linewidth.
To accelerate the data processing, we project the comb lines onto a two-dimensional CCD via a VIPA-grating system\cite{Diddams:07, BaoC:19}, enabling parallel processing on the comb lines.
Finally, a spectrometer with 200-kHz resolution, 4-THz bandwidth, and minutes-level processing time has been achieved.

\section*{Results}

\begin{figure*}[h!]
    \centering
    \includegraphics[width=0.95\linewidth]{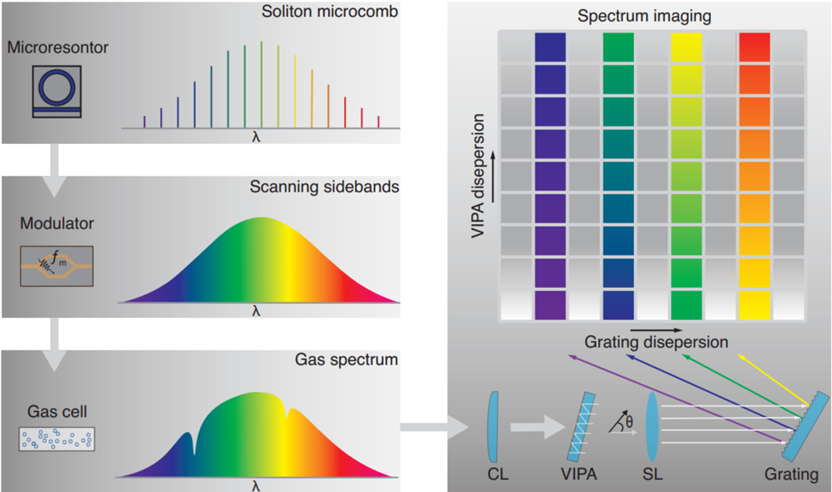}
    \caption{
    \scriptsize
    \textbf{Schematic and the principle of the microcomb spectrometer.}
    A soliton microcomb is generated in a Si$_3$N$_4$ microresonator.
    After passing through an electro-optic intensity modulator, the comb lines are suppressed, and symmetric sidebands are generated.
    By adjusting the modulation frequency, the sidebands can scan across the frequency gap between the comb lines.
    These scanning sidebands interact with matters, encoding their absorption spectral information.
    The absorption spectrum is then resolved using a VIPA-grating system.    
    CL: cylindrical lens; VIPA: virtually imaged phased array; SL: spherical lens.
    }
    \label{Fig:1}
    \end{figure*}

\noindent \textbf{Principle and experimental set-up}. 
The schematic of our microcomb spectrometer is illustrated in Fig. \ref{Fig:1}. 
A soliton microcomb generated in a Si$_3$N$_4$ microresonator serves as a broadband coherent light source with a repetition rate ${f_\mathrm{r}}=19.98$ GHz. 
However, directly using the microcomb for spectroscopy would limit its spectral resolution because it cannot sample spectral features between the comb lines \cite{suh:16}. 
To address this limitation, we employ an electro-optic intensity modulator to suppress each comb line carrier, generating two symmetric sidebands per comb line. 
The frequency difference between each sideband and its carrier is equal to the driven frequency ${f_\mathrm{m}}$.
By scanning ${f_\mathrm{m}}$, the sideband frequencies are adjusted accordingly. 
These scanned sidebands can interact with the device under test (DUT) to provide spectral information at frequencies between the comb lines. 
To resolve this spectral information, we use a VIPA-grating system, which maps the spectra onto a two-dimensional spatial plane captured by a CCD array. 

\begin{figure*}[h!]
    \centering
    \includegraphics[width=0.95\linewidth]{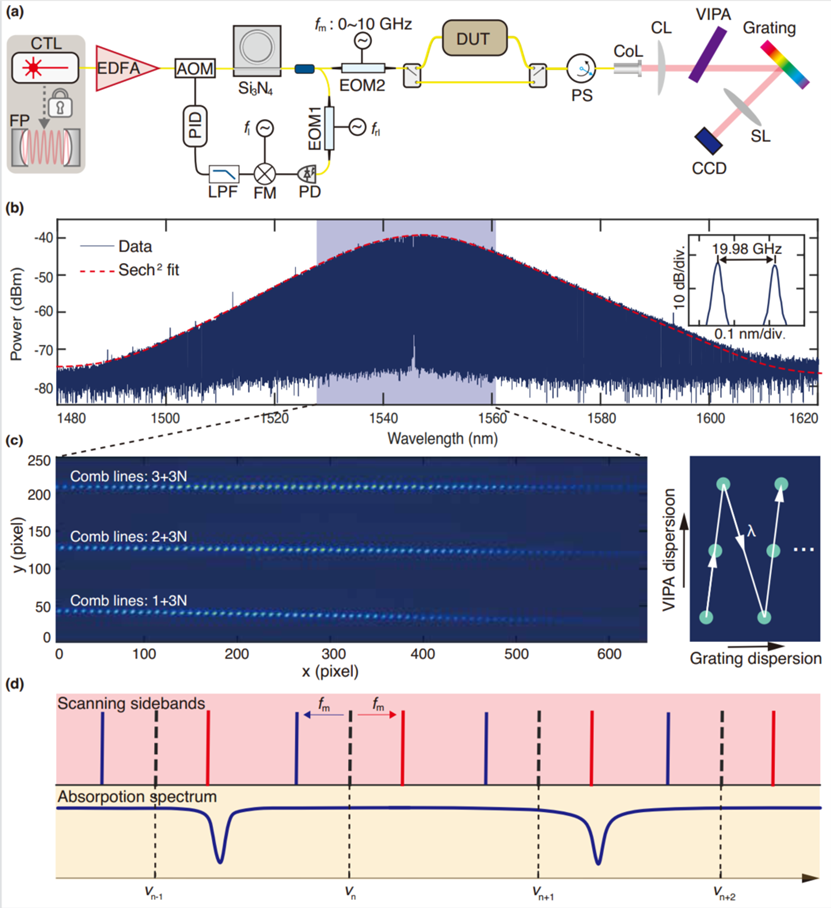}
    \caption{
    \scriptsize
    \textbf{Experimental setup and imaging process}.
    \textbf{(a)}. 
    Setup of the microcomb spectrometer. 
    CTL: continuously tunable laser; 
    FP: Fabry-Pérot cavity; 
    EDFA: erbium-doped fiber amplifier; 
    AOM: acoustic-optic modulator; 
    PID: proportional-integral-derivative controller; 
    LPF: low-pass filter; 
    FM: frequency mixer; 
    PD: photodetector; 
    EOM: electro-optic modulator; 
    DUT: device under test; 
    PS: polarization scrambler; 
    CoL: collimating lens. 
    \textbf{(b)}. 
    Spectrum of the microcomb. 
    The inset displays the zoomed-in diagram of two comb lines, with the mode spacing of 19.98 GHz.
    \textbf{(c)}. 
    % Imaging of the microcomb on the CCD Array.
    The left panel shows the imaging of the microcomb on the CCD.
    The right panel illustrates the dispersion directions of the grating and VIPA on the CCD. 
    The white arrows indicate the direction of wavelengths increasing in the image.
    \textbf{(d)}. 
    The principle for measuring spectrum using scanning sidebands.
    The dashed lines in the red zone indicate the locations of the comb lines before modulation.
    The blue and red lines represent the double sidebands when modulating the comb lines.
    $f_\mathrm{m}$ is the modulation frequency.
    The blue curve in the yellow zone is an absorption spectrum to be measured.}
    \label{Fig:2}
\end{figure*}

The experimental setup is depicted in Fig. \ref{Fig:2}(a).
A pump laser, tunable and amplified to ~800 mW via an erbium-doped fiber amplifier, is coupled into a Si$_3$N$_4$ microresonator to generate a soliton microcomb \cite{Guo:16}. 
The microcomb spectrum is shown in Fig. \ref{Fig:2}(b). 
Temperature control is used to stabilize the repetition rate ${f_\mathrm{r}}$, while the pump laser is locked to a reference FP cavity for long-term frequency stability (see Supplementary Note 1 for ${f_\mathrm{r}}$ characterization and Note 2 for pump laser locking characterization).

The microcomb is split into two parts. 
The weaker part enters an electro-optic modulator (EOM1) for light phase modulation driven at $({f_\mathrm{r}} - {f_\mathrm{l}})/2$, where ${f_\mathrm{l}}$ (traceable to a rubidium clock) serves as the locking reference \cite{Qing:24}. 
The modulated signal is detected by a high-speed photodetector (PD), mixed with ${f_\mathrm{l}}$, filtered by a low-pass filter (LPF), and fed into a proportional integral derivative (PID) controller. 
The output of the PID controller drives an acoustic-optic modulator (AOM) to lock ${f_\mathrm{r}}$ via controlling the microresonator pump power.

The stronger part enters another electro-optic modulator (EOM2) for intensity modulation, generating sidebands as shown in Fig. \ref{Fig:2}(d). 
The modulated microcomb is further split into two branches: one interacts with the DUT, while the other serves as a reference. 
Dual optical switches are utilized to minimize crosstalk-induced phase disturbances during laser selection and recombination. 

A high-speed polarization scrambler eliminates polarization sensitivity by randomizing the comb's polarization state. %before dispersion through a VIPA and a blazed grating.
The scrambled light is collimated into a 13-mm-diameter beam, focused by a cylindrical lens, and then dispersed by the VIPA and the blazed grating. 
As illustrated in Fig. \ref{Fig:1}, the VIPA projects the spectrum in the vertical direction.
While the VIPA provides superior spectral resolution, it suffers from light aliasing when the spectrum exceeds its FSR \cite{Xiao:04}. 
To mitigate this aliasing, the blazed grating further separates the spectrum along the horizontal axis, converting it into a two-dimensional spatial distribution.
An achromatic spherical lens focuses the resolved spectrum onto the CCD array with $ 1280\times1064$ pixels, the area of interest being shown in Fig. \ref{Fig:2}(c). 
Here, the VIPA has an FSR of 61 GHz, about three times the mode spacing of the soliton microcomb (19.98 GHz). 
Consequently, each comb line of the microcomb appears every three spots along the y-axis (vertical direction), resulting in three rows (horizontal direction) of well-ordered spots observed in the left image of Fig. \ref{Fig:2}(c).
The right panel of Fig. \ref{Fig:2}(c) illustrates the wavelength variation direction, with a white arrow indicating wavelength increasing. 
The CCD array then fetches the spectral data in a parallel way, which enables the rapid detection of the sidebands with broad spectral information and high resolution.

% Here, the VIPA has a thickness of 1.68 mm, a size of 1 inch, an FSR of 61 GHz, and a finesse of 71. 
% The blazed grating features a groove density of 600 lines/mm, a blaze wavelength of 1550 nm, a blaze angle of 26.78°, and a size of 1 inch.
% In this experiment, the microcomb has ${f_\mathrm{r}}=19.98$ GHz, and the VIPA's free spectral range (FSR) is 61 GHz.
% Consequently, each comb line of the microcomb appears every three spots along the y-axis, resulting in three rows of well-ordered spots observed in the left image of Fig. \ref{Fig:2}(c). 
% The right panel of Fig. \ref{Fig:2}(c) illustrates the wavelength variation direction, with a white arrow indicating increasing wavelength. 

\noindent \textbf{Spectral frequency determination}. 
% When the microcomb is modulated to generate sidebands, the CCD will record more complicated spots.
To determine the rough wavelength (or frequency) of each spot, a tunable single-frequency laser replaces the microcomb.
The laser is set to tune in steps of 0.02 nm. 
The CCD records the pixel position of each bright spot at different wavelengths, establishing a one-to-one correspondence between the pixel positions and the wavelengths. 
% This relationship enables interpolation to map wavelengths across the entire CCD pixel positions.
After the rough frequency calibration, the frequency of each comb line $\nu_n = \nu_\mathrm{p} \pm nf_\mathrm{r}$ can be roughly told from the CCD, where $\nu_\mathrm{p}$ is the pump frequency and $n$ represents a natural number (0,1,2,3...).
% The accurate frequency of each spot is then determined by calculating the sideband's frequency.
When modulation is applied, the frequency of the generated sideband can be expressed as $\nu_\mathrm{m} = \nu_n \pm f_\mathrm{m}$. 
Since $f_\mathrm{r}$ is locked and known, $f_\mathrm{m}$ is set by microwave source, $\nu_\mathrm{p}$ is measured by wavelength meter, and $n$ can be calculated by the rough frequency calibration of the CCD pixels position, each sideband's frequency is determined.
Scanning $f_\mathrm{m}$ from near zero to $f_\mathrm{r}/2 = 10$ GHz, the comb line spacing can be fulfilled by the sidebands.
This method enables laser frequency mapping across the entire CCD array.

It should be noted that, if $f_\mathrm{m}$ is too small, the two sidebands from the same comb line are too close to be distinguished on the CCD array. 
To address the issue, $f_\mathrm{m}$ is set to start at 1.5 GHz and increase up to 8.5 GHz with frequency step $f_\mathrm{step}$.
Then the microresonator's temperature is adjusted by 2 °C, which shifts the pump wavelength of the microcomb by approximately 4 GHz. 
After the pump wavelength is shifted, $f_\mathrm{m}$ is set to start at 1.5 GHz and increase up to 8.5 GHz again.
This shift allows the scanning sideband to cover the blind frequencies.

The VIPA-grating system provides parallel measurement on the information of the input spectrum.
The frequency accuracy is provided by the fully locked microcomb and the microwave-driven sidebands.
The sample resolution of our microcomb spectrometer is set by the sideband modulation frequency steps $f_\mathrm{step}$.
The ultimate spectral resolution is limited by the laser linewidth, which is 200 KHz in our system.

\noindent \textbf{Measurement of transmission spectra}.
We measured the spectra of three DUTs using our microcomb spectrometer. 
For HCN gas in a cell of 16-cm length and 25-torr pressure, the spectrum is measured with $f_\mathrm{step}=50$ MHz.
The measured HCN transmission spectrum from 1527.74 to 1560.63 nm is shown in Fig. \ref{Fig:3}(a) (see Supplementary Note 3 for HITRAN Data comparison).
Figure \ref{Fig:3}(b) details an individual line at 1554.56 nm fitted with a Voigt function, yielding a 2.00 GHz linewidth.
The measurement covers a spectral bandwidth of 4 THz because the spectrum beyond the 4 THz is not captured by the VIPA-grating system.
The whole measurement time takes approximately 3 seconds, which is constrained by the switching time of $f_\mathrm{m}$ and the CCD readout time.

Next, we measured the transmission spectrum of a Si$_3$N$_4$ microresonator with $f_\mathrm{step}=1$ MHz.
The transmission spectrum from 1544.33 to 1547.55 nm is illustrated in Fig. \ref{Fig:3}(c).
The measurement time is about 6 minutes.
Figure \ref{Fig:3}(d) shows a zoomed-in view of an absorption line at 1545.35 nm, fitted with a Lorentzian function, yielding a full-width at half-maximum (FWHM) linewidth of 45 MHz.

Lastly, a whispering-gallery-mode (WGM) resonator with an ultra-high $Q$-factor was measured with $f_\mathrm{step}=200$ kHz, corresponding to the laser linewidth.
The measured transmission spectrum from 1545.3779 to 1545.4001 nm is shown in Fig. \ref{Fig:3}(e).
The measurement time is about 6 minutes.
Figure \ref{Fig:3}(f) presents a resonance at 1545.3957 nm with a linewidth of 1.8 MHz.
These results demonstrated the sub-MHz resolution of our spectrometer.

\section*{Discussion}
\begin{figure*}[h!]
\centering
\includegraphics[width=0.95\linewidth]{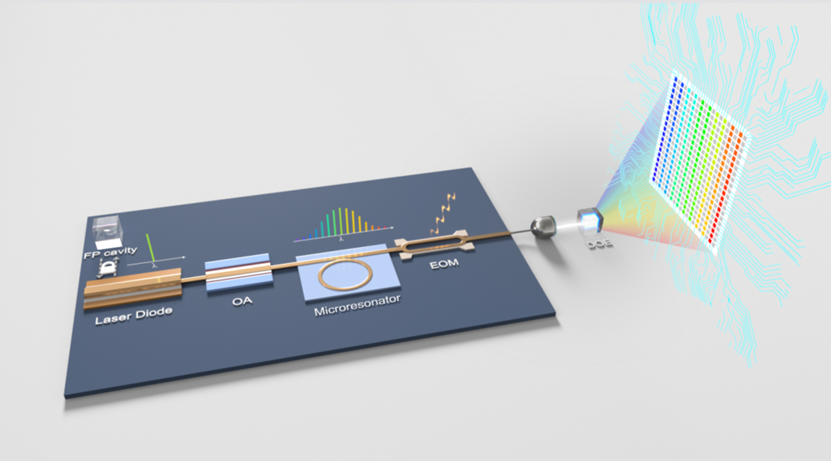}
\caption{
\scriptsize
\textbf{Envision of a miniaturized microcomb spectrometer}.
The core components of the microcomb spectrometer, including the laser source, the optical amplifier, the Si$_3$N$_4$ microresonator, and the electro-optic modulator, can be integrated monolithically on a chip. 
The others, such as the FP cavity, the DOE and the CCD, can be co-packaged with the chip.
OA: optical amplifier; DOE: diffractive optical element.
}
\label{Fig:4}
\end{figure*}

Crucially and promisingly, the system can be miniaturized with the advanced technologies of photonic integrated circuits (PICs) and photonic-electronic co-packaging\cite{Thomson:16, Rickman:14, Xiang:24}, as shown in Fig. \ref{Fig:4}.
Si$_3$N$_4$ waveguides fabricated by CMOS-foundry-compatible Damascene process\cite{Liu:21} are ideal for wafer bonding and micro-transfer printing\cite{Xiang:21,Chang:17,Roelkens:24}.
Lasers can be manufactured by III-V foundry\cite{Zhou:15, Zhou:23} and heterogeneously integrated with the Si$_3$N$_4$ waveguides for soliton microcomb generation\cite{Xiang:21}. 
% Photodetectors have been demonstrated in heterogeneous integration with the Si$_3$N$_4$ waveguides\cite{Yu:20}.
Silicon or LiNbO$_3$ EOMs can be integrated with Si$_3$N$_4$ waveguides\cite{Churaev:23}.
The optical amplifier can be the erbium-doped LiNbO$_3$\cite{Chen:21} or the III-V semiconductor optical amplifiers\cite{OpdeBeeck:21}.
Integrated ultrastable FP cavities have been demonstrated in Ref.\cite{Liu:24, Jin:22}.

In conclusion, we have demonstrated a soliton microcomb-based spectrometer with 4 THz bandwidth and 200 kHz resolution.  
Locking the pump laser to an FP cavity and referencing the repetition rate to a rubidium clock enabled fully locked microcomb generation. 
By combining the stable microcomb with intensity modulation, we transfer the sub-Hz precision of the radio frequency (RF) domain to the optical domain.
Further, sweeping the RF frequency, the modulated sidebands of the comb lines can fulfill the line gaps. 
Spectral analysis is performed using a VIPA and diffraction grating, which can project the modulated sidebands onto a CCD array for parallel data acquisition.  
The parallel measurement method allows broadband spectra analysis in minutes-level time.
With further refinements—such as employing an ultra-narrow-linewidth pump laser—Hz-level precision and resolution are attainable.
Therefore, our microcomb spectrometer releases the trade-offs among the wide bandwidth, the high resolution and high speed for spectroscopy for the first time, yet can be miniaturized by the mature CMOS and III-V foundries. 
This achievement paves the way for wide applications in sensing, communication, imaging, and quantum information processing.

%%%%%%%%%%%%%%%%%%%%%%%%%
%%%%%%%%%%%%%%%%%%%%%%%%%
%%%%%%%%%%%%%%%%%%%%%%%%%
%%%%%%%%%%%%%%%%%%%%%%%%%
%%%%%%%%%%%%%%%%%%%%
\begin{figure*}[h!]
    \centering
    \includegraphics[width=0.95\linewidth]{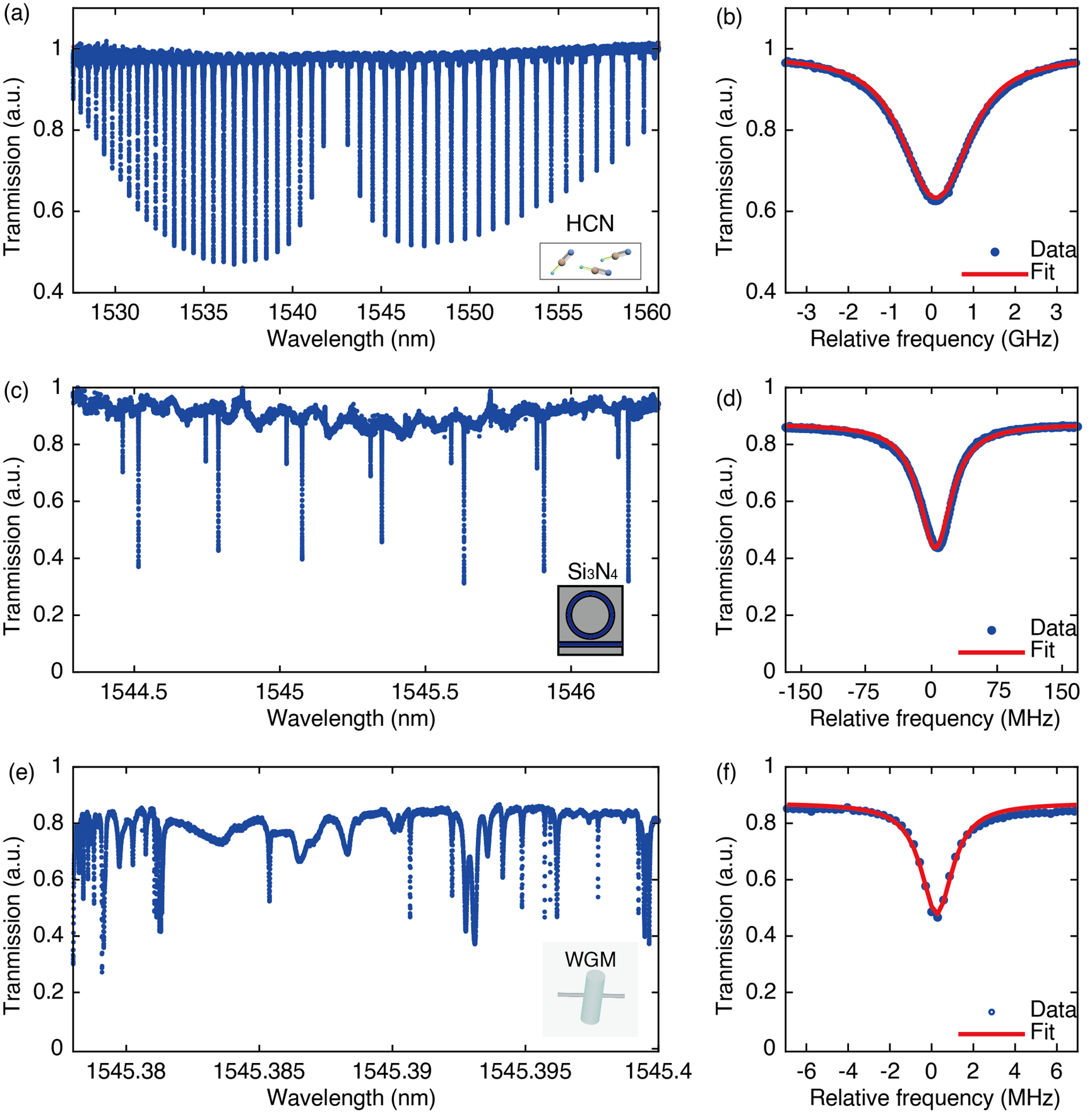}
    \caption{
    \scriptsize
    \textbf{Three applications on spectral analysis}.
    \textbf{(a)}. The spectral measurement of an HCN gas cell. 
    Panel \textbf{(b)} corresponds to the absorption line at 1554.56 nm in panel \textbf{(a)}, with the measured linewidth of 2.00 GHz. 
    \textbf{(c)}. The spectral measurement of a Si$_3$N$_4$ microresonator. 
    Panel \textbf{(d)} corresponds to the transmission absorption feature at 1545.35 nm in panel \textbf{(c)}, with the measured linewidth of 45 MHz.
    \textbf{(e)}. The spectral measurement of a WGM cylindrical cavity. 
    Panel \textbf{(f)} corresponds to the transmission absorption feature at 1545.3957 nm in panel \textbf{(e)}, with a linewidth of 1.8 MHz.
    }
    \label{Fig:3}
    \end{figure*}
%%%%%%%%%%%%%%%%%%%%
%%%%%%%%%%%%%%%%%%%%%%%%
%%%%%%%%%%%%%%%%%%%%%%%%%%
\clearpage\

%%%%%%%%%%%%%%%%%%%%%%%%%%%%%%%%%%%%%

%%%%%%%%%%%%%%%%%%%%%%%%%%%%%%%%%%%%%
\noindent \textbf{Acknowledgements}:This research was supported by National Natural Science Foundation of China (Grant No. 42125402, 42188101, 92476112, 42374185, 12261131503, 12404436); The B-type Strategic Priority Program of CAS (XDB0780000); Innovation Program for Quantum Science and Technology (2021ZD0300301, 2023ZD0301500); Guangdong-Hong Kong Technology Cooperation Funding Scheme (Grant No. 2024A0505040008); Shenzhen-Hong Kong Cooperation Zone for Technology and Innovation (HZQB-KCZYB2020050); Shenzhen Science and Technology Program (Grant No. RCJC20231211090042078); State Key Laboratory of Pulsed Power Laser Technology Foundation; Furthermore, ZRC is grateful for the support from USTC Tang Scholar; XXH is grateful for the support from the New Cornerstone Science Foundation through the XPLORER PRIZE.

\noindent \textbf{Author contributions}: X. X. supervised and directed the research. R. Z. and B. Y. jointly conceptualized the study and designed the experimental protocols. B. Y. led the execution of the experiments, with C. H. assisting in experimental operations and performing partial data processing. J. Li provided additional experimental support.The Si$_3$N$_4$ chip is fabricated by C. S. and J. Liu and characterized by B. S. and W. S.. C. W. contributed technical guidance on the operation of electronic analytical instrumentation. T. C. and C. L. coordinated laboratory resource management and equipment procurement. R. Z, X. X., and X. D. participated in manuscript quality control, critical discussion, and editorial refinement. All authors reviewed and approved the final version of the manuscript.

\noindent \textbf{Competing interests}: C. S. and J. Liu are co-founders of Qaleido Photonics, a start-up that is developing heterogeneous silicon nitride integrated photonics technologies. Others declare no competing interests.

\noindent \textbf{Data Availability Statement}: The data provided in the manuscript are available from the corresponding author upon reasonable request.
%%%%%%%%%%%%%%%%%%%%%%%%%%%%%%%%%%%%%

\clearpage
\end{document}

\title{Supplementary Information for: A high-resolution microresonator-frequency-comb spectrometer}

\author{Ruocan Zhao}
\thanks{These authors contributed equally to this work.}
\affiliation{CAS Key Laboratory of Geospace Environment, School of Earth and Space Sciences, University of Science and Technology of China, Hefei 230026, China}

\author{Bin Yang}
\thanks{These authors contributed equally to this work.}
\affiliation{CAS Key Laboratory of Geospace Environment, School of Earth and Space Sciences, University of Science and Technology of China, Hefei 230026, China}

\author{Chuan Huang}
\affiliation{CAS Key Laboratory of Geospace Environment, School of Earth and Space Sciences, University of Science and Technology of China, Hefei 230026, China}

\author{Jiangtao Li}
\affiliation{CAS Key Laboratory of Geospace Environment, School of Earth and Space Sciences, University of Science and Technology of China, Hefei 230026, China}

\author{Baoqi Shi}
\affiliation{International Quantum Academy, Shenzhen 518048, China}

\author{Wei Sun}
\affiliation{International Quantum Academy, Shenzhen 518048, China}

\author{Chen Shen}
\affiliation{International Quantum Academy, Shenzhen 518048, China}
\affiliation{Qaleido Photonics, Shenzhen 518048, China}

\author{Chong Wang}
\affiliation{CAS Key Laboratory of Geospace Environment, School of Earth and Space Sciences, University of Science and Technology of China, Hefei 230026, China}

\author{Tingdi Chen}
\affiliation{CAS Key Laboratory of Geospace Environment, School of Earth and Space Sciences, University of Science and Technology of China, Hefei 230026, China}
\affiliation{Hefei National Laboratory, University of Science and Technology of China, Hefei 230088, China}

\author{Chen Liang}
\affiliation{Hefei National Laboratory, University of Science and Technology of China, Hefei 230088, China}

\author{Xianghui Xue}
\email[]{xuexh@ustc.edu.cn}
\affiliation{CAS Key Laboratory of Geospace Environment, School of Earth and Space Sciences, University of Science and Technology of China, Hefei 230026, China}
\affiliation{Hefei National Laboratory, University of Science and Technology of China, Hefei 230088, China}
\affiliation{CAS Center for Excellence in Comparative Planetology, Anhui Mengcheng Geophysics National Observation and Research Station, University of Science and Technology of China, Hefei, China}
\affiliation{Hefei National Research Center for Physical Sciences at the Microscale and School of Physical Sciences, University of Science and Technology of China, Hefei, China.}

\author{Junqiu Liu}
\affiliation{International Quantum Academy, Shenzhen 518048, China}
\affiliation{Hefei National Laboratory, University of Science and Technology of China, Hefei 230088, China}

\author{Xiankang Dou}
\affiliation{Hefei National Laboratory, University of Science and Technology of China, Hefei 230088, China}

\maketitle
%\pagebreak

%%%%%%%%%%%%%%%%%%%%%%%%%%%%%%%%%%%%%%%%%%%%%%%%%%%%%%%%%%%%%%%
\section{Characterization of the microresonator and microcomb repetition rate stability test}
We used a vector spectrum analyzer \cite{Luo:24} to measure the transmission spectrum of the microresonator (TE mode) over the 1480-1640 nm wavelength range.
The dispersion of the microresonator is characterized by the integrated dispersion, defined as
\begin{linenomath*}
    \begin{equation}
    D_\text{int}(\mu)=\omega_\mu-\omega_0-D_1\mu=\sum_{n=2}^{\cdots}\frac{D_n\mu^n}{n!}
    \label{Eq.Dint}
    \end{equation}
\end{linenomath*}
where $\omega_{\mu}/2\pi$ is the $\mu^\text{th}$ resonance frequency relative to the reference resonance frequency $\omega_0/2\pi$,
$D_1/2\pi$ is microresonator FSR, 
$D_2/2\pi$ describes group velocity dispersion (GVD), 
and $D_n/2\pi~(n>2)$ are higher-order dispersion terms.
The $D_\text{int}$ profile is shown in Supplementary Fig. \ref{Fig:S1}(b).
A typical resonance and its fitting are presented in Supplementary Fig. \ref{Fig:S1}(c), showing an intrinsic linewidth of 20.5 MHz.
The experimental setup for soliton generation and soliton repetition rate characterization is illustrated in Supplementary Fig. \ref{Fig:S1}(a).
Supplementary Fig. \ref{Fig:S1}(d) displays the time-domain response of the microresonator as the pump light scans across the resonance at 800 mW power.
The soliton step is broadened to 200 MHz via an electro-optic modulation (EOM) for soliton addressing\cite{ZhengH:23}.
After locking the soliton repetition rate using the method described in the main text, we characterized it using an electrical spectrum analyzer.
The result, shown in Supplementary Fig. \ref{Fig:S1}(e), indicates a soliton repetition rate of 19.9759150 GHz with a standard deviation of 52 Hz.

\begin{figure}[b!]
\renewcommand{\figurename}{Supplementary Figure}
\centering
\includegraphics{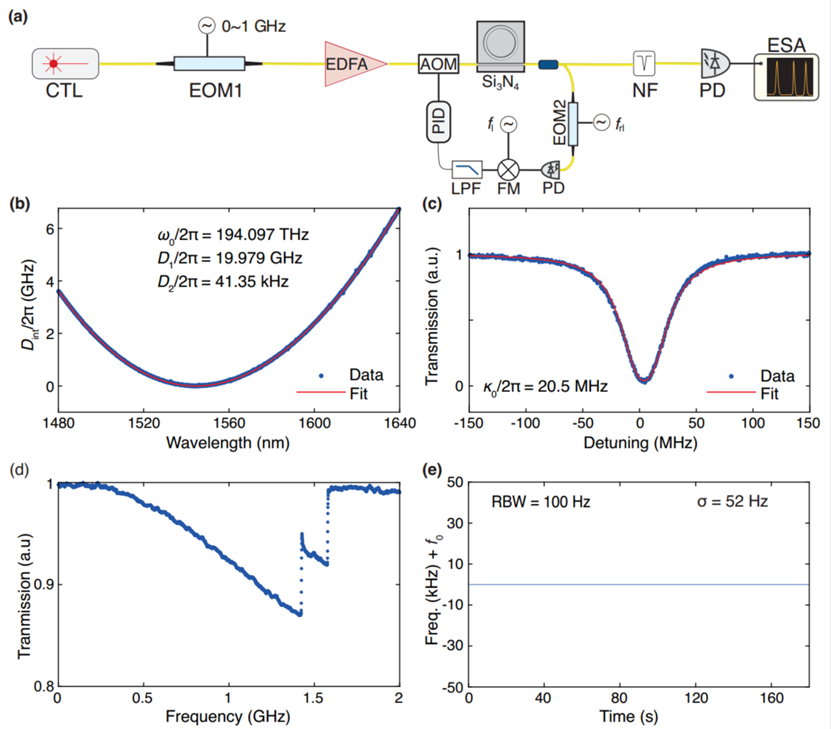}
\caption{
Microcomb repetition rate measurement and microresonator characterization. 
(a) Schematic of the microcomb repetition rate measurement. 
NF: notch filter; 
ESA: electrical spectrum analyzer. 
(b) Integrated dispersion profile of the microresonantor. 
(c) Transmission of a typical resonance. 
The intrinsic linewidth is fitted as $\kappa_0/2\pi = 20.5$ MHz.
(d) Time-domain response during the pump light scanning across a resonance from the blue-detuned side to the red-detuned side. 
(e) Microcomb repetition rate measurement results.
The standard deviation is measured as $\sigma = 52$ Hz.
$f_0 = 19.9759150$ GHz.
RBW: resolution bandwidth.
}
\label{Fig:S1}
\end{figure}
\clearpage

\section{Pump light locking}
%The spectral accuracy of the microcavity transmission spectrometer is not only dependent on the DKS repetition rate but also on the pump light. 
%For the DKS comb, the role of the pump light is similar to the initial frequency (carrier-envelope offset frequency) in a mode-locked comb, determining the overall offset of the DKS comb. 
%To improve the frequency accuracy of the DKS comb, although additional locking of the pump light is required, it also provides an absolute reference for the DKS comb. 
We use the Pound-Drever-Hall (PDH) technique to lock the pump light to an ultrastable Fabry-Pérot (FP) cavity.
The experimental setup is shown in Supplementary Fig. \ref{Fig:S2}(a).

The pump light is split into two paths.
In the first path, it is combined with a fiber optical frequency comb (OFC) locked to an ultrastable cavity and then filtered by a bandpass filter.
The beat signal is detected by a high-speed photodetector (PD) and sent to an electrical spectrum analyzer to monitor the stability of the pump light.

In the second path, the pump light is modulated by a phase electro-optic modulator (EOM) and then emitted into space.
It subsequently passes through a polarizing beam splitter (PBS) and enters the FP cavity.% to obtain its reflection absorption spectrum.
The reflected light is redirected by the PBS and captured by a high-speed PD.
The electrical signal from the reflected light is phase-demodulated to generate an error signal, which is fed back to the CTL laser via a PID controller for frequency compensation, achieving stable locking of the pump light.

Supplementary Fig. \ref{Fig:S2}(b) presents the beat frequency measurement between the pump light and the OFC after PDH locking.
The results show that the linewidth of the pump light is about 200 kHz.
Supplementary Fig. \ref{Fig:S2}(c) illustrates the long-term stability test of the pump light after PDH locking.

\begin{figure*}[h!]
\renewcommand{\figurename}{Supplementary Figure}
\centering
\includegraphics{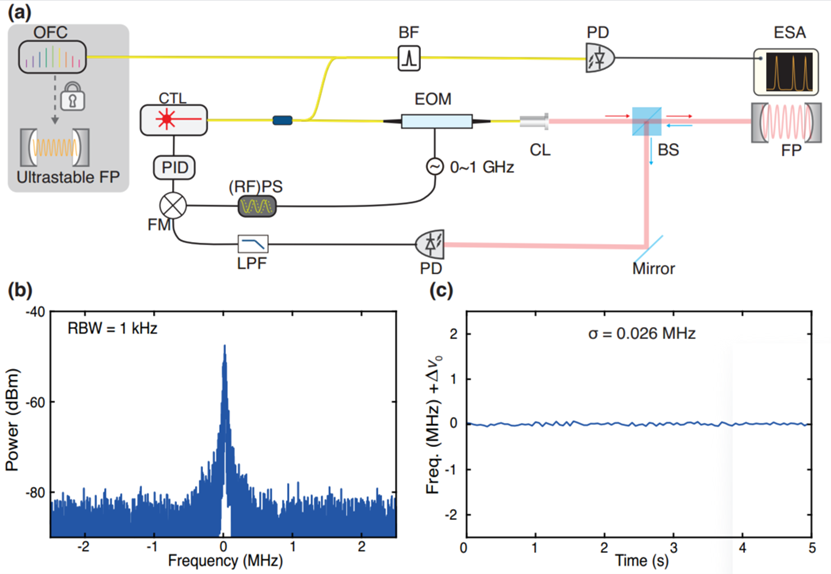}
\caption{Pump light locking and frequency characterization. 
(a) Schematic of the pump light locking. OFC: optical frequency comb; PBS: polarizing beam splitter; PS: phase shifter. 
(b) Beat frequency between the pump light after locking and the OFC. 
(c) Frequency stability of the pump light after locking.
The standard deviation is $\sigma = 0.026$ MHz.
$\Delta\nu_0 = 181.759$ MHz is a reference frequency.}
\label{Fig:S2}
\end{figure*}
\clearpage

\section{The transmission spectrum of the HCN gas}

\begin{figure*}[h!]
\renewcommand{\figurename}{Supplementary Figure}
\centering
\includegraphics{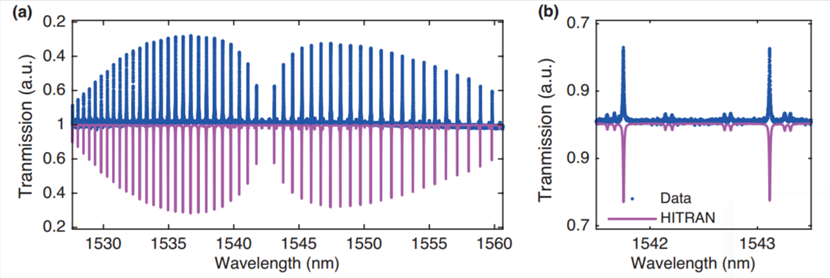}
\caption{%Comparison of the measured HCN gas transmission with the HITRAN database. 
(a) Direct comparison of the measured HCN gas transmission with the HITRAN database. 
(b) Zoomed-in view of the transmission spectrum around 1542 nm.}
\label{Fig:S3}
\end{figure*}

Supplementary Fig. \ref{Fig:S3} compares the HCN gas measurement results obtained using the microcomb spectrometer with the HITRAN database. 
Both the measured amplitude and the frequency by our spectrometer are agreeable with the HITRAN database.
% Supplementary Fig. \ref{Fig:S3}(b) presents a zoomed-in view of the transmission around 1542 nm.

%%%%%%%%%%%%%%%%%%%%%%%%%%%%%%%%%%%%%%%%%%%%%%%%%%%%%%%%%%%%
\vspace{1cm}
\section*{Supplementary References}